# A Summer Meridional Subsurface Temperature Dipole Mode in the South China Sea

**Ximing Wu[1], Fengchao Yao[1,2*], and Dongxiao Wang[1,2]**

[1] School of Marine Sciences, Sun Yat-sen University, Zhuhai, China
[2] Southern Marine Science and Engineering Guangdong Laboratory (Zhuhai), Zhuhai, China

Corresponding author: Fengchao Yao (yaofch@mail.sysu.edu.cn)

**Abstract**

The ocean heat content variability in the South China Sea (SCS) plays a pivotal role in regional climate and extreme weather events, such as tropical cyclones. Using high-resolution ocean reanalysis data, we show that the SCS exhibits a summer subsurface temperature dipole mode that controls the interannual variability of ocean heat content in the upper SCS. This dipole mode manifests as warm anomalies in the north and cold anomalies in the south during strong monsoon years, and a reversed pattern during weak monsoons years. The monsoon variability is linked to large-scale climate variability associated with El Niño–Southern Oscillation transitions. Heat budget analysis indicates that this dipole pattern is primarily driven by vertical heat transport linked to opposite wind stress curl anomalies in the northern and southern basin. Accompanying the vertical heat transports is a shallow meridional overturning circulation that redistributes heat between the northern and southern SCS.

**1. Introduction**

The ocean subsurface plays a critical role in regulating both oceanic and atmospheric processes and has attracted growing attention in recent years (Gao et al., 2025; He et al., 2024; Hu et al., 2021; Zhang et al., 2024). In particular, subsurface temperature and ocean heat content (OHC) are recognized as key factors influencing tropical cyclone (TC) genesis (Gao et al., 2022; Huang et al., 2015; Jin et al., 2014). As subsurface temperature accounts for a large portion of upper ocean's thermal conditions, understanding its variability is essential for improving climate prediction and weather forecast (Abraham et al., 2013; Cheng & Zhu, 2014; Xiao et al., 2019).

Unlike surface temperature, which is primarily influenced by atmospheric forcing, subsurface temperature is more strongly modulated by ocean dynamical processes (Song et al., 2014; Xiao et al., 2019; Yan et al., 2010). For example, mesoscale eddies can reshape the subsurface temperature through perturbed isotherms (G. Chen et al., 2012; Chen et al., 2011), and such eddy-induced anomalies are often more pronounced in the subsurface layer than the attenuated effects of surface thermal-forcing (Frenger et al., 2015; Zhang et al., 2016).

Wind stress curl (WSC) is another essential driver of subsurface temperature variability (Murphree et al., 2003; Sayantani & Gnanaseelan, 2015). Through Ekman suction and pumping, positive and negative WSC anomalies induced vertical displacement of isotherms and generate subsurface temperature anomalies (Ngo & Hsin, 2021; Yao & Wang, 2024). Seasonal variations in WSC have been shown to affect temperatures below the seasonal thermocline (Murphree et al., 2003). In the tropical Indian Ocean, for example, WSC anomalies drive downwelling and upwelling Rossby waves, inducing a north-south subsurface temperature dipole that is distinct from the conventional east-west dipole associated with the Indian Ocean Dipole (IOD) (Kakatkar et al., 2020; Sayantani & Gnanaseelan, 2015).

The South China Sea (SCS), the largest marginal sea in the western north Pacific, is frequently affected by the natural disasters such as TC and marine heatwaves during summer (Li et al., 2019; Wang et al., 2007; Yao & Wang, 2021). Both the frequency and intensity of these events exhibit significant interannual variability and an increasing trend (Song et al., 2023; Tan et al., 2022). Most previous studies on the temperature variability in the SCS have focused on the surface signals (Fang et al., 2006; C. Wang et al., 2006; Xiao et al., 2020; Yu et al., 2019), primarily due to the widespread availability of satellite-derived sea surface temperature (SST) data. However, emerging evidence suggests that the extreme events in the subsurface of the SCS are often more intense than those at the surface and do not show coherent variability with the surface (Yao & Wang, 2021, 2024; Zhang et al., 2025). Despite growing recognition of their importance, the spatiotemporal patterns of subsurface temperature in the SCS, and the underling mechanism remain largely unexplored.

During summer, the upper layer circulation of the SCS are primarily controlled by the southwesterly monsoons (Fang et al., 2002; Qu, 2000). The interaction between the monsoon and coastal mountains generates a distinct WSC structure over the SCS (Liu et al., 2008). The northern and southern SCS are governed by positive and negative WSC, respectively, with the strongest positive WSC occurring off the Vietnamese coast (11-14 °N) (Xie et al., 2003; Yu et al., 2020). The combined effects of the summer monsoon and the WSC give rise to an eddy pair off the Vietnamese coast, characterized by a strong cyclonic northern eddy and a weak anticyclonic southern eddy (C. Chen et al., 2012; Chu et al., 2017; Li et al., 2014; G. Wang et al., 2006; Xiang et al., 2016). The eddy pair has a marked impact on the thermohaline structure of the local upper ocean by inducing vertical transports (Chen et al., 2010; Ngo & Hsin, 2021). However, the eddy pair's influence is spatially limited to the Vietnamese coast.

In this study, we use long-term oceanic reanalysis data to analyze the interannual variability of subsurface temperatures in the SCS and unveil a dipole temperature mode that dominates variability across the basin-scale subsurface layer of the SCS. We further explore the underlying dynamical processes responsible for this mode and examine its relationship with large-scale climate variability associated with El Niño-Southern Oscillation (ENSO) transitions.

## 2. Data and Methods

### 2.1. Data

The 1/12° high-resolution temperature, current, and sea surface height (SSH) data from 1993 to 2018 for the analysis are derived from the Global Ocean Physics Reanalysis (GLORYS) product (Jean-Michel et al., 2021). Vertical velocity is diagnosed by integrating horizontal divergence (Johnson et al., 2001). To evaluate the reliability of the reanalysis data, we compare GLORYS with both observational and other reanalysis datasets. The observational reference is the monthly one-degree gridded Argo temperature dataset, which spans 2004-2022 and includes 58 vertical levels (Li et al., 2017). Because Argo coverage is sparse in the southern SCS, this comparison was limited to the northern region. GLORYS showed good agreement in this region (Supplementary Figure S1a). To validate the data over the entire SCS, especially the southern part, we also compared GLORYS with monthly temperature data from the Ocean Reanalysis System 5 (ORAS5) for 1993-2018 (Copernicus Climate Change Service, 2021). The results demonstrate that GLORYS effectively captures temperature variability at depth and agrees well with both Argo (in the north) and ORAS5 datasets (Supplementary Figure S1), with the dipole mode clearly observed in ORAS5-based products (Supplementary Figure S2). Monthly wind data and sea-level pressure data are derived from the fifth-generation European Centre for Medium-Range Weather Forecasts (ECMWF) atmospheric reanalysis of the global climate (ERA5) (Hersbach et al., 2020).

## 2.2. Methods

The wind stress curl, and Ekman pumping velocity are calculated as follows:

$$\text{Curl}\tau = \frac{\partial \tau_y}{\partial x} - \frac{\partial \tau_x}{\partial y}, \tag{1}$$

$$W_e = \nabla \times \left(\frac{\tau}{\rho_o f}\right), \tag{2}$$

where $\tau_x$ and $\tau_y$ denote the zonal and meridional components of surface wind stress, respectively. The wind stress curl ($\text{Curl}\tau$) is calculated from these two components. $W_e$ is the Ekman pumping velocity, $\rho_o$ is the reference density of seawater, and $f$ is the Coriolis parameter.

In this study, the governing equation for subsurface OHC is written as (Lee et al., 2004; Xiao et al., 2018):

$$\rho C_p \frac{\partial}{\partial t} \iiint T_{sub} dV = ADV_{vertical} + ADV_{horizontal} + R, \tag{3}$$

The terms in (3) represent, from left to right, subsurface OHC change, heat transport by horizontal advection, heat transport by vertical advection and the residual term. Here $\rho$ is a reference density for seawater (1027 kg m$^{-3}$); $C_p$ is the specific heat capacity of seawater (4007 J °C$^{-1}$ Kg$^{-1}$). $T_{sub}$ is the subsurface temperature. The residual term, R, accounts for unresolved processes, such as diffusion and sub-grid-scale mixing (Xiao et al., 2018). $ADV_{vertical}$ and $ADV_{horizontal}$ are calculated as follows:

$$ADV_{horizontal} = \rho C_p \int \int v_S(T_S - T_{ref})dxdz$$
$$-\rho C_p \int \int v_N(T_N - T_{ref})dxdz$$
$$+\rho C_p \int \int u_W(T_W - T_{ref})dydz \quad (4)$$
$$-\rho C_p \int \int u_E(T_E - T_{ref})dydz,$$

$$ADV_{vertical} = \rho C_p \int \int w_{Bottom}(T_{Bottom} - T_{ref})dxdy$$
$$-\rho C_p \int \int w_{top}(T_{top} - T_{ref})dxdy, \quad (5)$$

where the subscripts N, S, W, E, top, and bottom represent the north, south, west, east, top, and bottom boundaries, respectively. T is the seawater temperature, and $T_{ref}$ is the reference temperature, which is calculated as the volume-mean temperature.

## 3. Results

### 3.1. Interannual Variability of the Subsurface Temperature in the SCS

3.1.1. Leading Mode of the Subsurface Temperature Variability in the SCS

The spatiotemporal characteristics of summer subsurface temperature in the SCS were examined using Empirical Orthogonal Function (EOF) analysis of temperature averaged between 50 m and 150 m (Figure 1a). The leading mode (EOF1), accounting for about 33% of the total variance, reveals a pronounced meridional dipole pattern, with a boundary near 11°N. The corresponding principal component (PC) time series is strongly correlated (r = 0.69) with the intensity of the SCS summer monsoon, defined as the wind stress averaged over the SCS basin (1–18°N, 104–118°E)(Wu et al., 2025) (Figure 1b). Composite analyses of subsurface (50-150 m) temperature anomalies during strong (1994, 2002, and 2018) and weak (1995, 1998, and 2010) summer monsoon years—defined as years when summer wind intensity exceeds the 90th percentile or falls below the 10th percentile from 1993 to 2018—further confirm the presence of the dipole mode, with opposite polarities under strong and weak monsoon conditions (Figures 1c and 1d). For reference, the spatial distribution of subsurface temperature anomalies for individual anomalous years from 1993 to 2018 is shown in Supplementary Figure S3, along with the composite results based on ORAS5 data (Figure S2), which also consistently exhibit the dipole pattern.

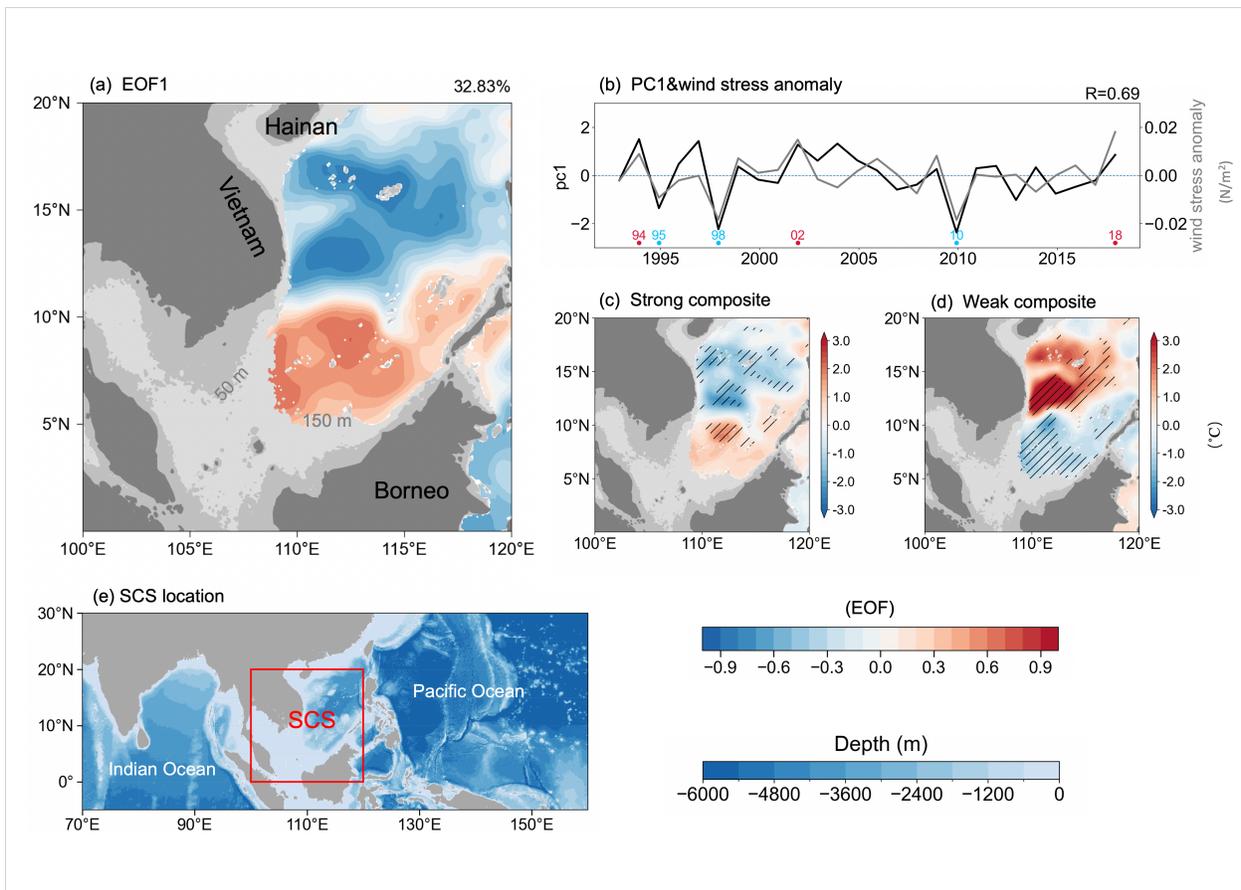

**Figure 1. A dipole mode that dominates summer subsurface temperature variability in the SCS.** (a) The first mode of Empirical Orthogonal Function (EOF) analysis of August subsurface temperature (50–150m) in the SCS. The 50 m and 150 m

depth contours are shown in light grey. August is chosen instead of the summer mean for the subsurface temperature in this study to avoid potential data biases introduced during the averaging process (Chu et al., 2017). (b) The principal component (PC) time series corresponding to the first EOF mode and the time series of SCS summer mean (July, August, and September) monsoon intensity anomaly (N/m$^2$), averaged over the region (1–18°N, 104–118°E). Red and blue dots represent years of strong and weak summer monsoon, respectively. The relative threshold for anomalous summer monsoon is defined as the wind stress intensity exceeding the 90th percentile of its summer values over the period 1993-2018 (Wu et al., 2025). (c) and (d) Composite anomalies of August subsurface temperature in the SCS under strong and weak summer monsoon conditions, respectively. Hatched areas indicate regions that are statistically significant at the 95% confidence level. (e) Regional context map showing the location of the SCS (red box) within the tropical Indo-Pacific.

3.1.2. Vertical Structure of the Subsurface Temperature Meridional Dipole

To better understand the horizontal and vertical structure of the subsurface dipole mode, we focus on two representative years: the strong summer monsoon of 2018 and the weak summer monsoon of 2010, both of which clearly exhibit the dipole mode (Figures 2 and S3).

In 2018, negative subsurface temperature anomalies occupied the northern SCS (> 11°N) while positive anomalies dominated the southern SCS (< 10 °N). This dipole-like temperature structure spanned the basin-wide subsurface layer, with the most pronounced anomalies occurring between 50 and 150 m, reaching maximum depths of approximately 300 m (Figures 2a and 2d). Subsurface cooling co-occurred with negative SSH anomalies, whereas subsurface warming was associated with positive SSH anomalies, indicating a strong connection between upper-layer dynamics and subsurface temperature anomalies. Notably, the surface layer did not exhibit a dipole pattern. Instead, it showed basin-wide cooling due to the combined effect of stronger evaporative cooling and enhanced upwelling (Xie et al., 2003).

During the weak monsoon year of 2010, the dipole mode reversed polarity: warm anomalies developed in the northern SCS and cool anomalies appeared in the south (Figure 2b). The vertical structure closely resembled that of 2018, with subsurface temperature anomalies again closely aligned with SSH variations. Similar to 2018, the surface layer lacked a dipole mode but instead exhibited basin-scale warming due to the combined effect of reduced evaporation and weakened upwelling (C. Wang et al., 2006; Xie et al., 2003; Yao & Wang, 2021). In the following analysis, we show that the subsurface dipole structure is more related to the dynamical response of the SCS to the variability of the summer monsoon.

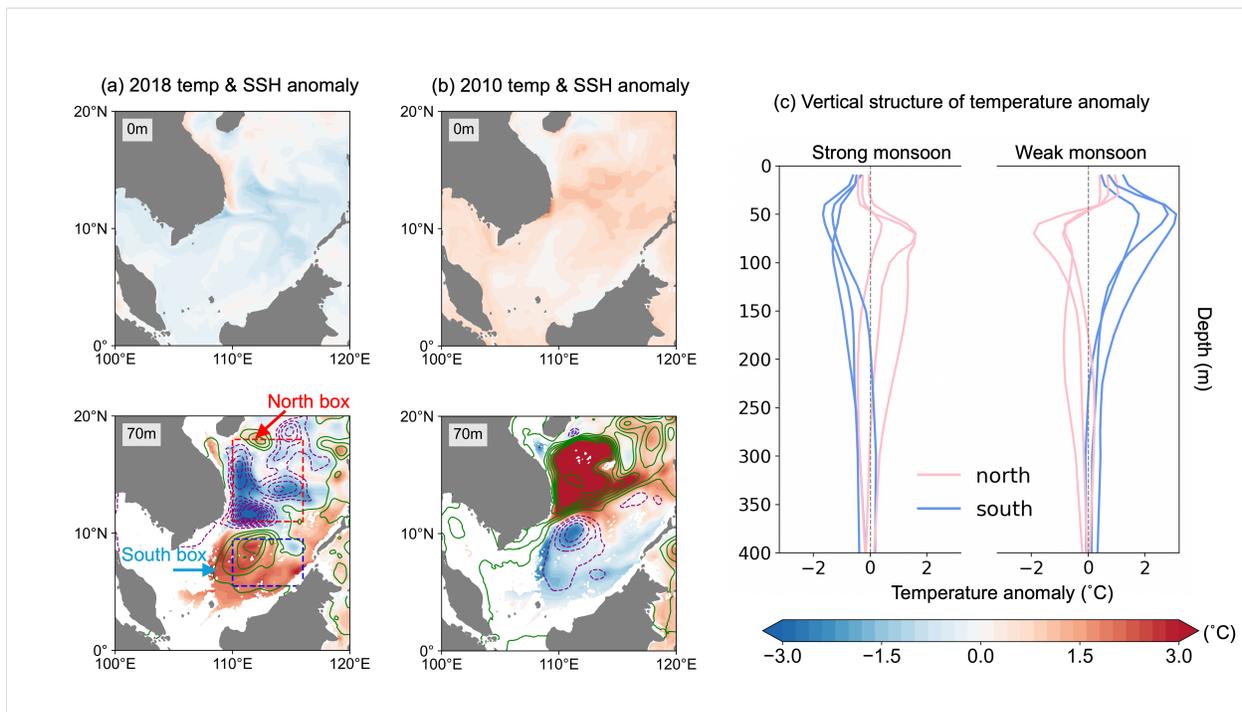

**Figure 2. Vertical and horizontal temperature structure of the SCS under anomalous summer monsoon conditions.** (a) Temperature anomalies (°C) at various depths in the SCS under the strong summer monsoon in August 2018. (b) same as (a), but for August 2010. Solid green (dashed purple) contours denote positive (negative) SSH anomalies, with a contour interval of 2.5 cm. (c) Vertical structure of temperature anomalies in the northern (red box in (a); 110-116°E, 11-18°N) and southern (blue box in (a); 110-116°E, 5.5-9.5°N) SCS in August during strong and weak summer monsoon years.

3.1.3. Links between the Dipole Mode and Large-Scale Climate Variability

The intensity of the SCS summer monsoon, which significantly influences the subsurface temperature dipole pattern, is closely linked to large-scale climate variability. The monsoon intensity tends to strengthen during the developing phase of El Niño and weaken during its decaying phase (Chang et al., 2008; Li et al., 2014). This variation is due to the temperature gradient between the Tropical Indian Ocean (TIO)-SCS region and the equatorial Pacific Ocean, which then modulates the SCS summer monsoon (Chen et al., 2016; Chu et al., 2017).

As shown in Figure 3a, during strong summer monsoon years, the Pacific exhibits an El Niño-like pattern, corresponding to the developing phase of El Niño. In this phase, the central and eastern equatorial Pacific show positive SST anomalies, while the TIO and the SCS experience negative SST anomalies. This reversed inter-basin temperature gradient between the TIO-SCS cooling and the equatorial Pacific warming generates a cyclonic low sea-level pressure anomaly over the northwestern Pacific, which enhances SCS summer monsoon (Wu et al., 2025). The enhanced summer monsoon, combined with coastal mountains, leads to a southwesterly wind stress anomaly over the SCS, with the strongest values located around 11°N. This wind stress pattern further results in positive WSC anomalies in the northern SCS and negative WSC anomalies in the southern SCS (Figure 3c).

Conversely, during weak summer monsoon conditions, corresponding to the decaying phase of El Niño, the pattern is reversed, leading to a weakened SCS summer monsoon (Figure 3b). In particular, during the rapid transition from El Niño to La Niña, TIO warming acts like a capacitor, helping the anomalous anticyclone over the tropical northwest Pacific to persist from winter to summer, thereby weakening the SCS summer monsoon (Xie et al., 2009; Xie et al., 2016). Under the weakened summer monsoon, the wind stress anomaly over the SCS manifests primarily as a northeasterly jet south of 11°N. This anomaly causes a WSC anomaly pattern reversed from the strong monsoon conditions, with negative anomalies in the northern SCS and positive anomalies in the southern SCS (Figure 3c).

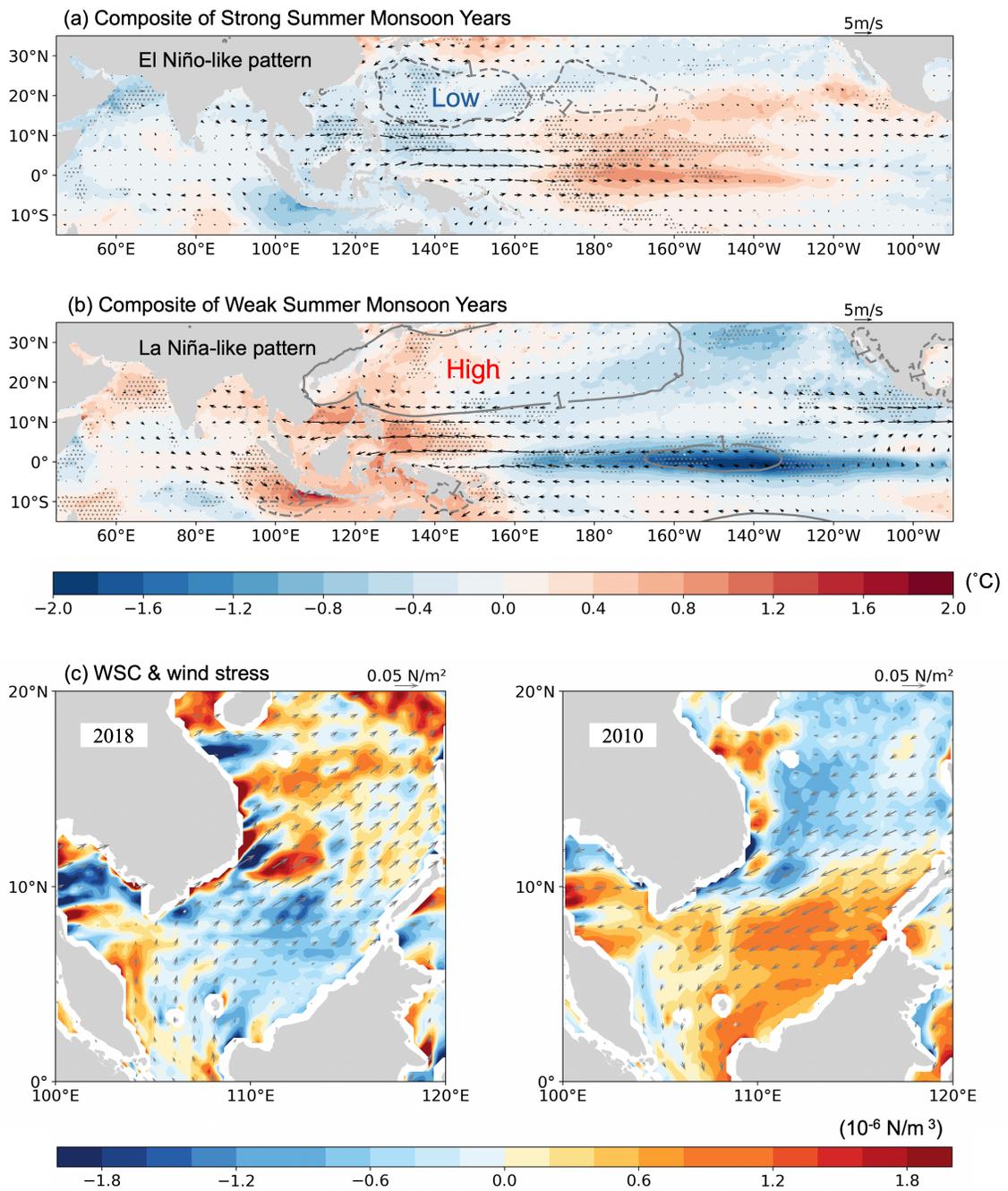

**Figure 3. Anomalous climate forcing associated with strong and weak SCS summer monsoon conditions.** (a) Composite map of large-scale summer mean (July, August, and September) anomalies for the 850 hPa wind (vectors, m/s), SST (colors, °C), and sea-level pressure (contours, hPa) under strong summer monsoon years. Stippling indicates areas where SST are statistically significant at the 95% confidence level. (b) Same as (a), but for weak summer monsoon years. (c) Summer mean (July, August,

and September) WSC anomalies (colors, $10^{-6}$ N/m$^3$) and surface wind stress (vectors, N/m$^2$) anomalies in 2018 and 2010, respectively.

## 3.2. Dynamical Processes Controlling the Subsurface Dipole Temperature Anomalies

3.2.1. Heat Budget and Associated Physical Processes

We next examine the processes responsible for the formation of the subsurface dipole mode and the associated physical drivers. The variation of total vertical velocity in the SCS is significantly correlated with Ekman pumping velocity: the correlation coefficient between observed anomalous Ekman pumping velocity and vertical velocity from GLORYS reanalysis reaches 0.78 in the northern SCS (red box in Figure 2a) and 0.71 in the southern SCS (blue box in Figure 2a) for the upper 50 m layer, and 0.76 and 0.56, respectively, for the upper 150 m (Supplementary Figure S4). This indicates that the Ekman pumping velocity effectively captures the primary variation of total vertical velocity in the SCS. To further quantify the contribution of wind-driven processes, we calculate the ratio of Ekman pumping velocity to total vertical velocity. Results show that Ekman pumping velocity accounts for approximately 46% of the total vertical velocity in the northern SCS and 67% in the southern SCS during summer. The lower value in the north is attributed to the influence of coastal upwelling off northern Vietnam, which has been previously shown to be weaker than the Ekman pumping effect but still contributes to the total vertical motion (Yu et al., 2020). Overall, these findings suggest that wind-driven processes dominate the vertical motion in this region. In the following section, we explore how the associated physical processes influence subsurface temperature variability.

To quantify the relative importance of the various driving factors, we examine the subsurface OHC budget (50-150 m) for the northern and southern SCS during 2010 and 2018. Figure 4a and 4b show an analysis of the subsurface OHC anomalies, decomposed into anomalies in the subsurface OHC tendency, alongside horizontal and vertical advection terms relative to 1993-2018 climatological mean. The subsurface OHC budget reveals that the subsurface temperature anomalies in both northern and southern SCS are because of vertical advection, whereas the horizontal advection plays a relatively minor role and, in some cases—such as the southern box in 2018 and both boxes in 2010—partially offsets the effects of vertical advection (Figure 4b). These results suggest that the summer subsurface dipole-like temperature anomalies in the SCS are primarily driven by vertical heat transport under anomalous monsoon conditions. The following will provide a detailed description of the processes underlying this anomalous vertical heat transport, followed by a discussion of the role of horizontal advection.

3.2.2. Strong Summer Monsoon

During the strong summer monsoon year of 2018, a subsurface cold anomaly developed in the northern SCS, while a warm anomaly emerged in the southern SCS. We analyzed the summer mean (July-September) of each term in the OHC budget anomaly for the northern and southern subsurface (50-150 m) temperatures. Results show that in the northern SCS, vertical advection term contributes approximately $-1.9 \times 10^{13}$ W and dominates the OHC tendency term. Horizontal heat advection terms play relatively weaker roles, with opposing effects across the southern boundary partially offsetting each other. The dominant vertical processes are closely linked to the WSC anomalies over the SCS. Specifically, positive WSC anomalies over the northern SCS induces Ekman suction, which shoals the isotherms and brings colder water upwards, thereby resulting in negative temperature anomalies (Figure 4c). This process is reflected at the surface as negative SSH anomalies (Figure 2a and 4c), corresponding to cyclonic eddy anomalies (Chen et al., 2010; Wu & Chang, 2005). It is observed that the northern SCS is covered with relatively small negative SSH anomalies (Figure 2a), which align closely with regions of negative subsurface temperature anomalies, corresponding to the widespread negative subsurface temperature anomalies across the northern SCS.

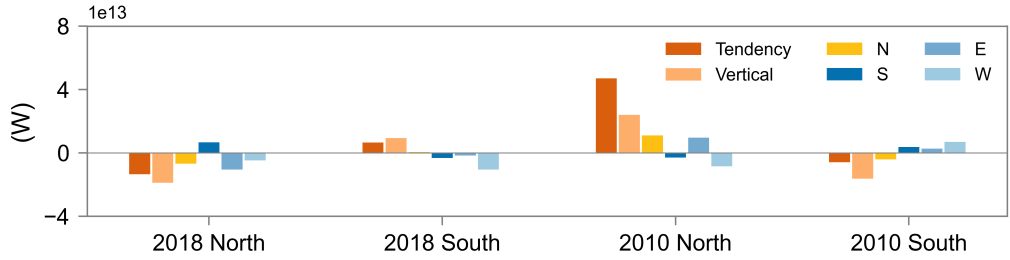

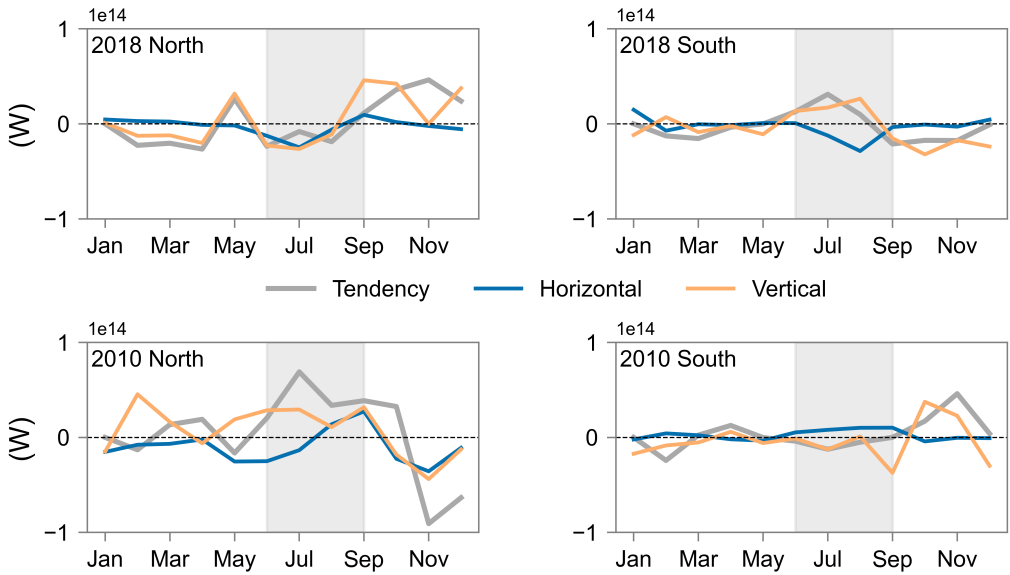

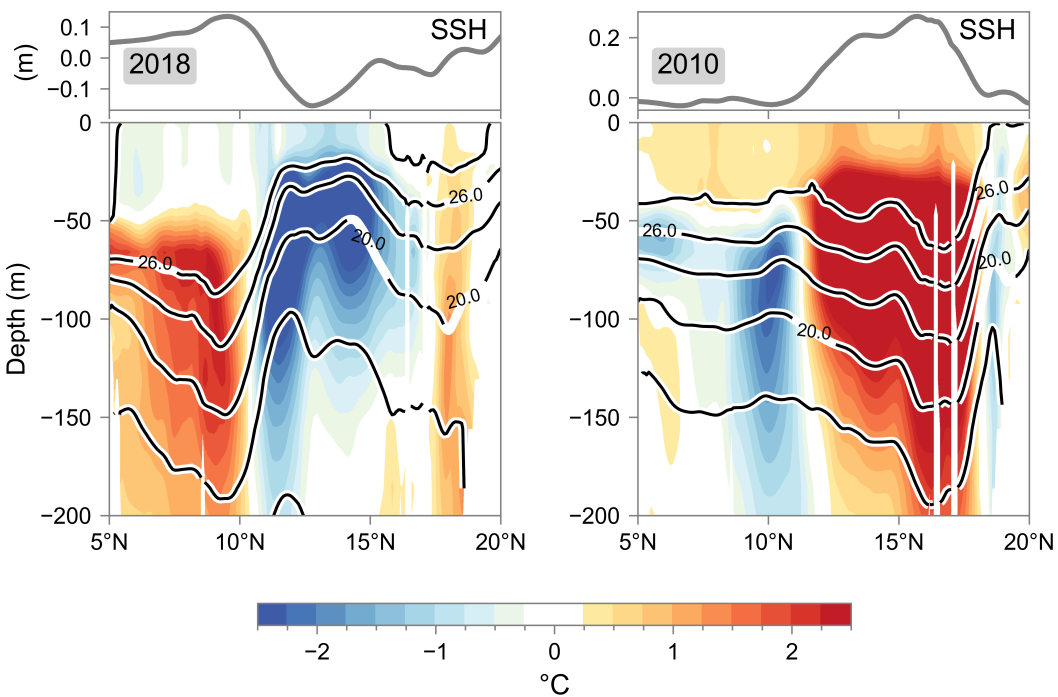
17

**Figure 4. Dynamical processes driving subsurface temperature anomalies.** (a) Anomalous subsurface OHC (50-150m) in the southern (red box in Figure 2a) and northern SCS (blue box in Figure 2a) for 2010 and 2018, respectively, decomposed into vertical advection term and each horizontal advection term. All terms represent the July-September mean. The N, S, E, W denote the horizontal advection across the northern, southern, eastern, and western boundaries, respectively. (b) Anomalous subsurface OHC tendency (50-150m) over the southern and northern SCS for the same period, further decomposed into vertical advection term and total horizontal term. (c) Cross-sections of SSH (m) and Temperature anomalies (°C) along 111.5°E (5-20°N) for August 2018 and 2010, respectively, illustrating the vertical structure associated with the dipole mode.

In the southern SCS, vertical advection term once again plays a dominant role, contributing approximately $0.9 \times 10^{13}$ W to the positive subsurface temperature anomalies. This vertical advection results from negative WSC, which induce downward Ekman pumping. As a result, the isotherms deepen, and warm surface waters are transported downward into the subsurface layer, resulting in a positive subsurface temperature anomaly. At the surface, this process is manifested as positive SSH anomalies (Figure 2a), corresponding to an anticyclonic circulation anomaly. Distinct from the northern SCS, the southern SCS subsurface is largely dominated by a single large-scale anomalous anticyclonic gyre, which leads to the formation of a broad positive SSH anomaly at the surface and subsurface temperature anomaly in the southern SCS.

3.2.3. Weak Summer Monsoon

Under weak summer monsoon conditions, the dipole pattern reverses, with warm subsurface anomalies emerging in the northern SCS and cold anomalies in the southern SCS. This reversal is attributed to the change in the sign of WSC anomalies over the SCS. In the northern SCS, the negative WSC anomalies lead to the formation of positive SSH anomalies, suppressing upwelling and facilitating downwelling (Yao & Wang, 2024). This leads to the downward bending of isotherms, which forces warm surface waters downward into the subsurface, resulting in positive subsurface temperature anomalies (Figure 4c). In contrast, in the southern SCS, positive WSC anomalies generate negative SSH anomalies, further weakening or eliminating downwelling and preventing warm surface waters from being transported into the subsurface, leading to negative subsurface temperature anomalies.

### 3.3. Meridional Overturning Circulation Associated with Temperature Anomalies

The subsurface temperature dipole mode identified here profoundly influences the meridional distribution of OHC within the upper 150 m of the SCS, which implies a dynamic process through which heat is redistributed across the basin meridionally. For instance, during the strong summer monsoon of 2018, warm subsurface temperature anomalies in the southern SCS and cold anomalies in the north led to an increase in upper 150 m OHC by approximately $1.33 \times 10^{20}$ J in the south, while the north experienced a decrease of about $-2.62 \times 10^{20}$ J. In contrast, during the weak summer of 2010, the dipole reversed, resulting in an increase in upper 150 m OHC by roughly $6.53 \times 10^{20}$ J in the north and a decrease of about $-3.48 \times 10^{19}$ J in the south. Supplementary Figure S5 shows the corresponding OHC anomalies for 2010 and 2018, as well as for the strong and weak monsoon composites.

By calculating the zonal average of meridional velocity across the SCS basin, we find that these vertical motions are accompanied by opposing horizontal advection anomalies in the surface and subsurface layers, which contribute to the development of a shallow meridional overturning circulation and compensate for the upper-layer OHC anomalies.

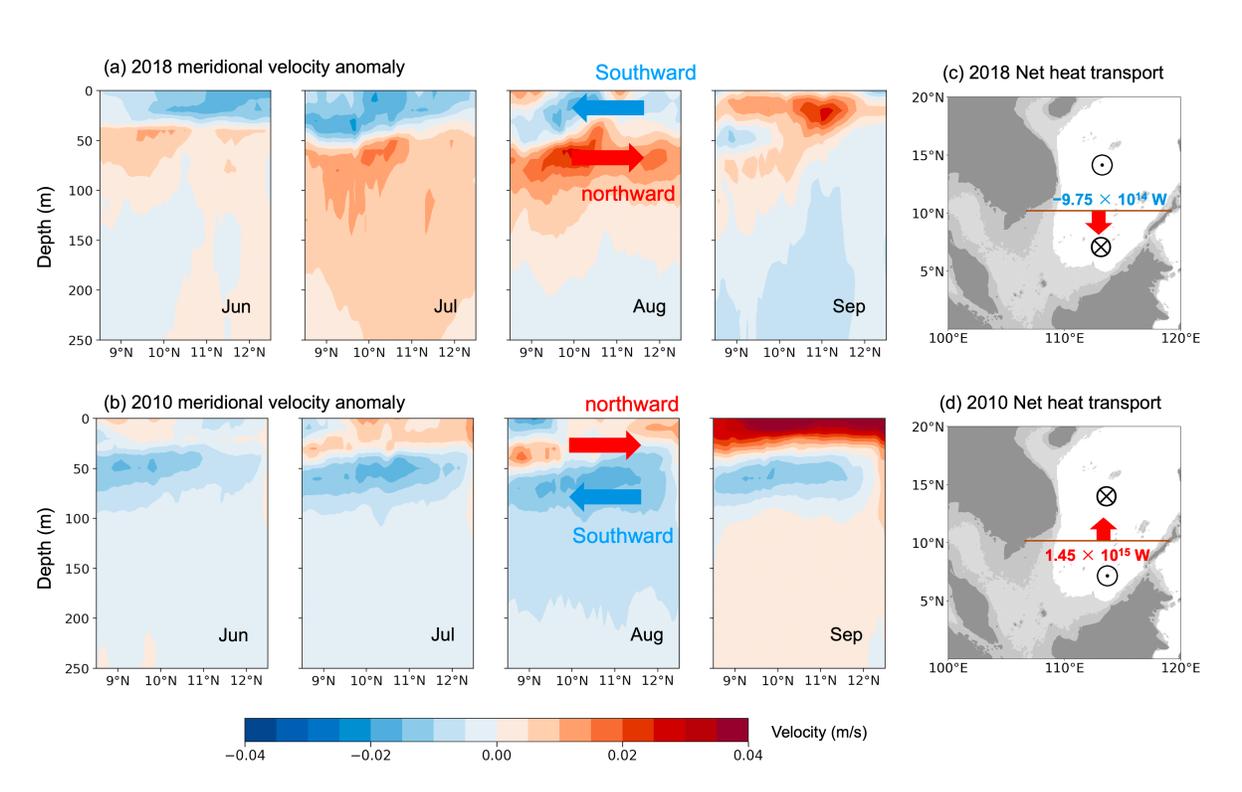

**Figure 5. Meridional overturning circulation associated with the anomalous dipole mode.** Panels (a) and (b) show meridional velocity anomalies for 2018 and 2010, respectively, zonally averaged across the basin (units: m/s). Panels (c) and (d) display the net heat transport in the upper 150 m across a meridional section at 10.5°N, between the southern and northern boxes, associated with these meridional velocity anomalies. The dotted circles represent upwelling, and the crossed circles indicate downwelling. A composite analysis for strong and weak summer monsoon years is provided in the Supplementary Figure S6 and further supports the presence of the meridional overturning circulation.

During the strong summer monsoon of 2018, the meridional velocity zonally averaged across the basin reveals a southward velocity anomaly in the 0-50 m layer, and a northward velocity anomaly in the 50-150 m layer (Figure 5a). These opposing flows generate a shallow, counterclockwise meridional overturning circulation, with downwelling in the south and upwelling in the north. Composite results for all strong summer monsoon years further confirm these opposing velocity anomalies between the surface and subsurface layers, supporting the existence of the counterclockwise overturning circulation (Supplementary Figure S6).

To quantify the heat transport by the meridional overturning circulation, we calculate the upper 150 m heat flux across a meridional section at 10.5°N between the southern and northern boxes (Figure 5c). The results indicate that this meridional overturning circulation transports heat southward, with a net southward heat transport anomaly of $-9.75 \times 10^{14}$ W in the upper 150 m, which contributes to the observed north-south contrast in upper-layer OHC.

In contrast, during 2010, the 0–50 m layer exhibited a northward velocity anomaly, while the 50–150 m layer showed a southward velocity anomaly (Figure 5b). These opposing velocity anomalies generated a clockwise meridional overturning circulation, featuring upwelling in the south and downwelling in the north. This pattern is also supported by composite results for all weak summer monsoon years (Supplementary Figure S6). The anomalous overturning circulation redistributed the upper OHC within the SCS, resulting in a net northward heat transport of $1.45 \times 10^{15}$ W in the upper 150 m (Figure 5d).

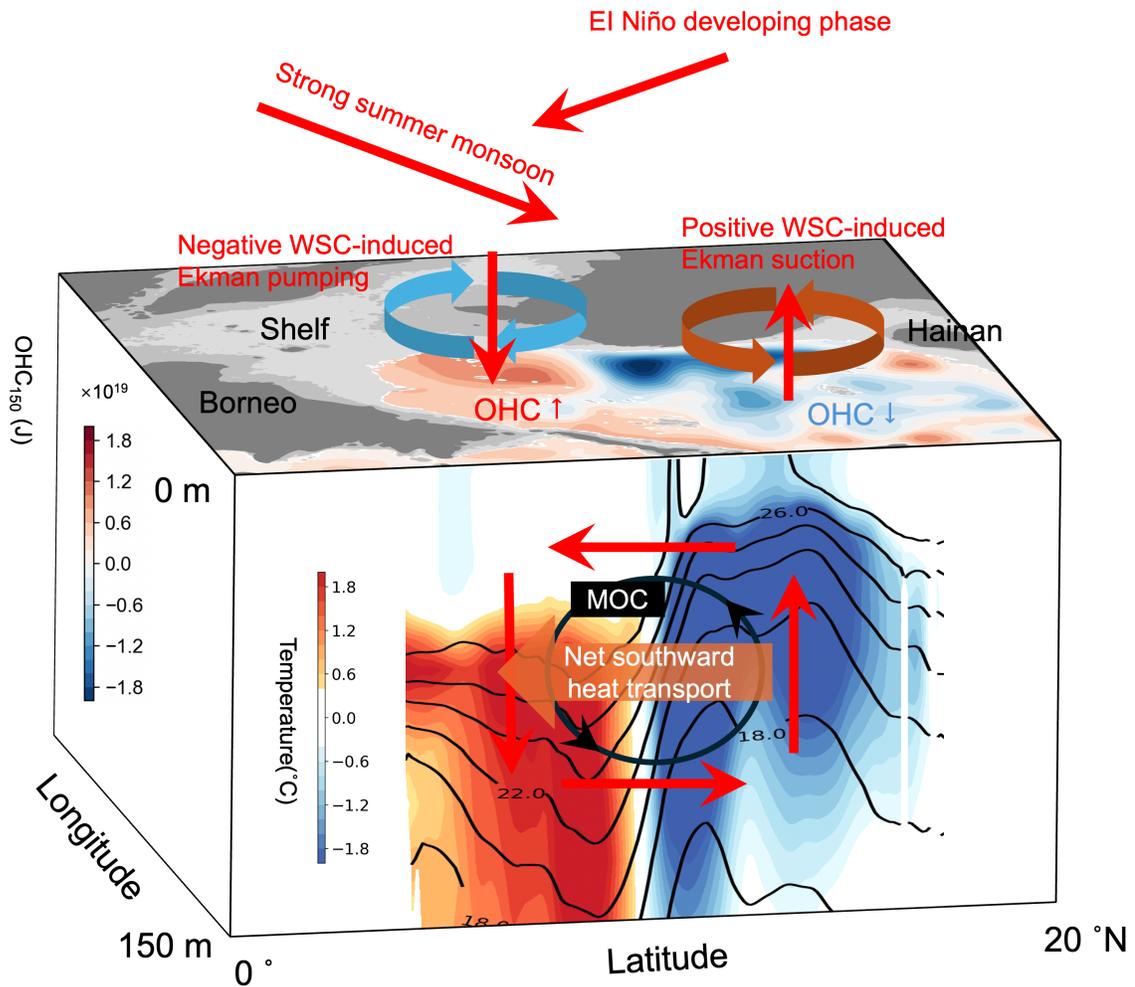

**Figure 6. Schematic illustrating the physical mechanisms driving the subsurface dipole mode in the SCS.** During the developing phase of El Niño, a strong southwesterly summer monsoon anomaly combines with the orographic effect of the Vietnamese coastal mountains, generating negative WSC anomalies in the southern SCS and positive WSC anomalies in the northern SCS. These WSC anomalies drive Ekman pumping in the south and Ekman suction in the north, resulting in positive subsurface temperature anomalies in the southern SCS and negative anomalies in the north. The resulting anomalous vertical motions are accompanied by a meridional overturning circulation that transports surface warm water southward and subsurface cold water northward, resulting in a net southward heat transport. This overturning circulation helps to compensate for the north-south upper-layer OHC contrast initially caused by subsurface temperature dipole. The horizontal panel shows upper 150 m OHC

anomaly, while the vertical panel depicts the temperature anomaly structure along 111.5°E. MOC refers to the meridional overturning circulation.

## 4. Conclusions

Using GLORYS data, this study reveals that the primary interannual variability of the subsurface temperature in the SCS exhibits a meridional dipole mode. Results of the subsurface OHC governing equation indicate that this pattern is primarily controlled by distinct vertical transport driven by WSC anomalies. The subsurface dipole mode further leads to the contrasting north-south contrast in upper layer OHC. The source of this heat contrast is attributed to the meridional overturning circulation identified in this study, which, through opposing horizontal flows in the upper and lower branches, transports heat between the southern and northern SCS, thereby modulating the upper layer OHC (Figure 6).

During strong summer monsoon in the developing phase of El Niño, the southern and northern SCS are characterized by the positive and negative subsurface temperature anomalies, respectively. The subsurface OHC controlling equation indicates that these anomalies are primarily driven by vertical transport due to distinct WSC anomalies above the SCS. Positive WSC anomalies in the northern SCS raise the isotherms and transport cold water upward, leading to negative subsurface temperature anomalies. Negative WSC anomalies in the southern SCS depress the isotherms and push the warm water downward, causing positive subsurface temperature anomalies. The subsurface dipole mode is accompanied by a counterclockwise meridional overturning circulation that transports surface warm water southward and subsurface cold water northward, resulting in a net southward heat transport. This overturning circulation helps to compensate for the meridional upper-layer OHC contrast initially caused by subsurface temperature dipole.

During weak summer monsoon conditions in the decaying phase of El Niño, the anomalies reverse, with negative subsurface temperature anomalies in the southern SCS and positive anomalies in the northern SCS. These anomalies remain primarily driven by opposing WSC anomalies. Negative WSC anomalies in the northern SCS suppress upwelling or even induce downwelling, leading to positive subsurface temperature anomalies. Conversely, positive WSC anomalies in the southern SCS weaken or eliminate downwelling, resulting in negative subsurface temperature anomalies. The reversed dipole mode is accompanied by a clockwise meridional overturning, resulting a net northward heat transport.

The meridional subsurface dipole mode identified in this study, characterized by its large basin-wide scale and persistence throughout the summer months, may exert a significant influence on the dynamics of subsurface marine heatwaves (He et al., 2024; Yao & Wang, 2024). Furthermore, the study underscores the connection between subsurface temperature variability in the South China Sea (SCS) and large-scale climate phenomena such as ENSO, suggesting a new pathway through which ENSO may affect the genesis and development of tropical cyclones (TCs). Given that TCs are highly sensitive to subsurface heat content in the SCS (Gao et al., 2025; Huang et al., 2015), this finding has important implications for understanding and predicting TC activity.

**Acknowledgments**

**Data availability statement**

The high-resolution temperature and current data from 1993 to 2020 are derived from the Global Ocean Physics Reanalysis (GLORYS) product at https://data.marine.copernicus.eu/product/GLO

BAL_MULTIYEAR_PHY_001_030/services. The monthly one-degree gridded Argo temperature data are available at https://argo.ucsd.edu/data/argo-data-products/. The ORAS5 temperature data are available from https://cds.climate.copernicus.eu/datasets/reanalysis-oras5?tab=download. The ERA5 wind data and sea-level pressure data are available from https://www.ecmwf.int/en/forecasts/dataset/ecmwf-reanalysis-v5. Nino 3.4 index is available at https://www.ncei.noaa.gov/access/monitoring/enso/sst.